\documentclass[%
 reprint,
 superscriptaddress,
nofootinbib,
 amsmath,amssymb,
 aps,
 prc,
]{revtex4-2}

\usepackage{multirow}
\usepackage{tabularx}
\usepackage{xcolor}
 \usepackage[section]{placeins} 
 
\definecolor{britishracinggreen}{rgb}{0.0, 0.26, 0.15}
\usepackage{cancel}

\definecolor{forestgreen}{rgb}{0.13, 0.55, 0.13}

\newcommand{\trento}{\texttt{T$_\mathrm{R}$ENTo}}
\newcommand{\tre}{\texttt{T$_\mathrm{R}$E}}

\newcommand{\SMASH}{\texttt{SMASH}}

\newcommand{\music}{\texttt{MUSIC}}
\newcommand{\is}{\texttt{iS3D}}

\newcommand{\distas}[1]{\mathbin{\overset{#1}{\; \sim}}}%
\usepackage{comment}
\usepackage{graphicx, url}% Include figure files
\usepackage{hyperref}
\graphicspath{ {./AUAUGrad/}{./PbPbCE/}{./PbPbPTB/}}
\usepackage{dcolumn}% Align table columns on decimal point
\usepackage{bm}% bold math
\begin{document}
\preprint{APS/123-QED}

\title{Efficient emulation of relativistic heavy ion collisions with transfer learning}

\author{D.~Liyanage}
\affiliation{Department of Physics, The Ohio State University, Columbus OH 43210.}

\author{Y.~Ji}
\affiliation{Department of Statistical Science, Duke University, Durham NC 27708.}

\author{D.~Everett}
\affiliation{Department of Physics, The Ohio State University, Columbus OH 43210.}

\author{M.~Heffernan}
\affiliation{Department of Physics, McGill University, Montr\'{e}al QC H3A\,2T8, Canada.}

\author{U.~Heinz}
\affiliation{Department of Physics, The Ohio State University, Columbus OH 43210.}
 
\author{S.~Mak}
\affiliation{Department of Statistical Science, Duke University, Durham NC 27708.}

\author{J.-F.~Paquet}
\affiliation{Department of Physics, Duke University, Durham NC 27708.}
 
%

\begin{abstract}

Measurements from the Large Hadron Collider (LHC) and the Relativistic Heavy Ion Collider (RHIC) can be used to study the properties of quark-gluon plasma. Systematic constraints on these properties must combine measurements from different collision systems and methodically account for experimental and theoretical uncertainties. Such studies require a vast number of costly numerical simulations. While computationally inexpensive surrogate models (``emulators'') can be used to efficiently approximate the predictions of heavy ion simulations across a broad range of model parameters, training a reliable emulator remains a computationally expensive task. We use transfer learning to map the parameter dependencies of one model emulator onto another, leveraging similarities between different simulations of heavy ion collisions. By limiting the need for large numbers of simulations to only one of the emulators, this technique reduces the numerical cost of comprehensive uncertainty quantification when studying multiple collision systems and exploring different models.
\end{abstract}

%\keywords{Suggested keywords}%Use showkeys class option if keyword display desired}

\maketitle

%%%%%%%%%%%%%%%%%%%%%%%%%%%%%%%%%%%%%%%%%%%%%%%%%%%%%%%%%%%%%%%%%%%%
\section{Introduction}
\label{sec1}
\vspace*{-2mm}
%%%%%%%%%%%%%%%%%%%%%%%%%%%%%%%%%%%%%%%%%%%%%%%%%%%%%%%%%%%%%%%%%%%%

The RHIC and LHC collider facilities create nuclear matter under extreme conditions by colliding heavy nuclei at relativistic velocities. These high energy collisions melt the nuclei and create a strongly interacting, exotic phase of nuclear matter called quark-gluon plasma (QGP) \cite{Gyulassy:2004zy}. The QGP filled the universe microseconds after the Big Bang, before it cooled down to produce atomic hydrogen, helium and other light atomic nuclei that we observe in the universe today \cite{Yagi:2005yb}. Due to its extremely short lifetime (${\sim\,}10^{-23}$\,s) and size (${\sim\,}10^{-14}$\,m), the QGP created in relativistic heavy ion collisions cannot be observed directly; it can only be studied through the final particles it emits.  

Modeling of relativistic nuclear collisions is a challenge that involves a succession of phases of many-body nuclear physics with different degrees of freedom; the QGP is only one of them. Realistic numerical simulations of such collisions have many physical parameters that are related to the properties of this QGP. To constrain these properties, one must effectively solve the inverse problem, i.e. find the model parameters, including their uncertainties, for which simulated observables agree well with the experimental data. 

Relativistic heavy ion collision experiments have accumulated a vast body of measurements and are continuing to do so. These experimental data vary widely in the size of their uncertainties, which can also have non-trivial correlations.  Theoretical simulations add additional uncertainties to the error budget, of two different types: statistical (aleatoric) uncertainties from measuring a finite number of samples from a stochastic process, and systematic (epistemic) uncertainties arising from imperfect modeling of the (not yet fully understood or only approximately implemented) physics underlying the dynamical evolution process. These experimental and theoretical uncertainties limit the precision with which the desired model parameters can be inferred.

Bayesian inference or Bayesian parameter estimation is a modern statistical method that provides a way to reliably infer the properties of QGP, by accounting methodically for both theoretical and experimental uncertainties. Tremendous progress has been made in the study of relativistic heavy ion collisions over the past decade by providing increasingly reliable constraints and error estimates for the properties of QGP using Bayesian statistical techniques \cite{Petersen:2010zt, Novak:2013bqa, Sangaline:2015isa, Bernhard:2015hxa, Bernhard:2016tnd, Moreland:2018gsh, Bernhard:2018hnz, Bernhard:2019bmu, JETSCAPE:2020shq, Nijs:2020ors, Nijs:2020roc, JETSCAPE:2020mzn}. As both the model and data have uncertainties, comparing them results in a probability distribution for the model parameters, specifying the probability for a model with a given set of parameters to provide predictions that agree with the experimental observations. A single model with $n$ parameters will have an $n$-dimensional probability distribution, called in brief ``the posterior'', describing its agreement with a set of measurements. For a class of competing models, the dimensionality of model parameter space increases accordingly. Bayesian uncertainty quantification depends on the ability to accurately sample this posterior probability distribution, which is generally not known analytically \cite{Trotta_2008}. Markov Chain Monte Carlo (MCMC) techniques provide such sampling methods \cite{MCMC}. They are practical only if fast approximations of otherwise expensive computer simulations are available. Emulation with surrogate models has thus become an essential component in any Bayesian inference involving a computationally expensive likelihood function. 

Emulators are machine learning models that provide a computationally efficient prediction of the simulator over the parameter space when trained on a sparse set of full simulation data. While a modeler can choose from a wide range of learning models (e.g., linear regression, decision trees, neural networks) as surrogates for expensive simulations, the standard practice in relativistic nuclear physics \cite{Novak:2013bqa, Sangaline:2015isa, Bernhard:2015hxa, Bernhard:2016tnd, Moreland:2018gsh, Bernhard:2018hnz, Bernhard:2019bmu, JETSCAPE:2020shq, Nijs:2020ors, Nijs:2020roc, JETSCAPE:2020mzn} has been to use Gaussian Process (GP) emulators \cite{santner2003design}. There are two reasons for this: (i) GPs provide a flexible non-parametric framework for emulation modeling and (ii) they also provide an efficient quantification of the predictive uncertainty associated with the interpolation between training points in the $n$-dimensional parameter space. In Bayesian parameter estimation, the latter integrates seamlessly with the aleatoric and epistemic uncertainties to yield an accurate quantification of the total uncertainty for the inferred model parameters. 

Relativistic heavy ion collision experiments have been conducted at various experimental facilities around the world, using different collision systems (ranging from p+p and p+$A$ to U+U) and different  collision energies (ranging from $\sqrt{s_\mathrm{NN}}=3$\,GeV to 13\,TeV).\footnote{%
    For readers trying to follow this rapidly-evolving field we recommend the series of proceedings for the annual to biannual {\it Quark Matter} conferences, the latest of which is published in \cite{QM2019, QM2019URL}.}
When studying these different systems with Bayesian parameter inference methods, one typically builds separate emulators for each individual system. Each collision is simulated using a multistage model \cite{Bass:2000ib, Nonaka:2006yn, Hirano:2007ei, Petersen:2008dd, Song:2010aq, Heinz:2011kt, Song:2013qma, Zhu:2015dfa, Ryu:2017qzn, Gale, JETSCAPE:2020mzn} that describes the successive dynamical evolution stages. For each stage there typically exist multiple physics models (``modules'') based on different physics assumptions. Mixing-and-matching these modules leads to a plethora of theoretical models that, in principle, could all be used to simulate the collision. As recently shown using Bayesian Model Averaging \cite{JETSCAPE:2020shq}, this ambiguity in the theoretical framework can add a significant model uncertainty in the parameter inference. But accounting for it systematically requires studying multiple models, and this generates a need for efficient emulators describing the predictions from different but typically closely related evolution models. If each model emulator needs the same number of training data, the computational cost for building the emulators  scales linearly with the number of models. This quickly renders a global Bayesian parameter inference, which includes a representative set of simulation models to describe large sets of experimental data from a variety of collision systems, computationally infeasible.
 
We introduce here a novel emulation method that significantly reduces the computational barrier for a global Bayesian parameter estimation by requiring a smaller volume of training data for building accurate emulators. This is accomplished by realizing that physical observables from different collision systems are related to each other by common trends resulting from the uniqueness of the underlying physics, and that predictions for these observables from models based on different sets of approximations for this underlying physics also share common trends reflecting this common ancestry. We use ``transfer learning" \cite{BMC, kennedy2000predicting, surveyTL} to transfer knowledge about such trends from emulators for a specific model trained on a larger, much more expensive set of already existing training data generated for a previously analyzed system, to new emulators for a different simulation model of the same collision system or for simulations of a different collision system. We provide illustrative examples on the use of this new technique; the code\footnote{%
    We use the EMUKIT package \cite{emukit2019} to implement transfer learning emulation.} 
generating these examples, including full documentation, can be found at \cite{gitcode}.

This work is organized as follows. Sec.~\ref{sec:transfer_learning} provides an introduction to transfer learning and Gaussian Process emulation. Applications of transfer learning techniques for emulation of relativistic heavy ion collisions are introduced and illustrated in Sec.~\ref{sec3}. In Sec.~\ref{sec:sens} we illustrate a new way of performing sensitivity analysis offered by transfer learning. We then compare the accuracy of and computational savings from the new emulation method to the existing usage of Gaussian Processes in Sec.~\ref{sec5}. Applications of this method and its limitations in analyzing relativistic heavy ion collisions and beyond are discussed in Sec.~\ref{sec6}. We conclude in Sec.~\ref{sec7} with an outlook on future work. The Appendix describes the standardization process for experimental observables used in our work.

%%%%%%%%%%%%%%%%%%%%%%%%%%%%%%%%%%%%%%%%%%%%%%%%%
%%%%%%%%%%%%%%%%%%%%%%%%%%%%%%%%%%%%%%%%%%%%%%%%%
\vspace*{-2mm}
\section{Transfer Learning and Gaussian Process Emulation}
\label{sec:transfer_learning}
\vspace*{-2mm}

%%%%%%%%%%%%%%%%%%%%%%%%%%%%%%%%%%%%%%%%%%%%%%%%%
%%%%%%%%%%%%%%%%%%%%%%%%%%%%%%%%%%%%%%%%%%%%%%%%%

%%%%%%%%%%%%%%%%%%%%%%%%%%%%%%%%%%%%%%%%%%%%%%%%%
\subsection{Transfer learning}
\label{sec2a}
\vspace*{-2mm}
%%%%%%%%%%%%%%%%%%%%%%%%%%%%%%%%%%%%%%%%%%%%%%%%%

Transfer learning methods (see, e.g., \citep{surveyTL, torrey2010transfer}) aim to improve learning in a designated task (called the \textit{target} task), by leveraging information from other related tasks (called \textit{source} tasks). This is in contrast to traditional machine learning methods, which instead build separate learning models for each task in isolation. Transfer learning methods are becoming increasingly popular in the machine learning literature, since it allows for efficient learning of target systems where training data can be expensive to obtain \cite{shavlik}.

While there are many types of transfer learning models, the one most relevant for the current study is \textit{inductive} transfer learning \citep{surveyTL}, where the source and target problems have identical input domains but different tasks. In such problems, the training data for the target task is typically scarce, so a model trained solely on such data does not provide good predictive performance. Existing transfer learning techniques tackle this problem by learning and correcting the bias between source and target tasks. One such method is TrAdaBoost \citep{TrAdaBoost}, which weighs each source data point by a measure of similarity to the target for better classification performance on the target task. This approach is extended for regression tasks in \cite{BoostingForRegression}. \cite{weightITL} proposes an importance-weighted approach for reweighing the source data to predict on the target task. The authors of \cite{cao2010adaptive} present an adaptive transfer learning model using Gaussian processes, in which a transfer kernel learns to model similarities between target and source tasks. Their model assumes the same kernel for both target and source, with a dissimilarity parameter accounting for the correlation between them. Our proposed model builds on these ideas but takes instead an additive approach where we introduce a discrepancy function between source and target, modeled by a GP. This provides a more flexible way of transferring information and also makes it possible to analyze the differences between source and target via sensitivity analysis on the discrepancy function. A comprehensive survey on existing transfer learning techniques can be found in \cite{zhuang2020comprehensive}.

The proposed transfer learning emulator is based on the popular Kennedy-O'Hagan (KO) model for multi-fidelity emulation \cite{kennedy2000predicting}. Here we address the bias between target and source by applying a correlation factor and a discrepancy function. This work provides a novel application of the KO model for modeling heavy ion collisions between different nuclear species, or for the same species using different but related dynamical evolution codes.

%%%%%%%%%%%%%%%%%%%%%%%%%%%%%%%%%%%%%%%%%%%%%%%%%
\vspace*{-2mm}
\subsection{Gaussian process emulation}
\label{sec2b}
\vspace*{-2mm}
%%%%%%%%%%%%%%%%%%%%%%%%%%%%%%%%%%%%%%%%%%%%%%%%%

Gaussian processes (GPs) \cite{Rasmussen2004} are a popular choice for emulation of computer simulations \cite{GP_sacks} and have been exploited in diverse applications from rocket design \cite{mak2018efficient} to 3D printing \cite{chen2019function}. GPs are an essential tool for Bayesian parameter estimation of complex simulation models, where they are used to efficiently interpolate between full model runs taken on a sparse set of design points in a high-dimensional parameter space, largely due to their ability to efficiently provide a probabilistic quantification of the incurred interpolation uncertainty.

Let $f(\mathbf{x})$ denote the simulation output at parameter point $\mathbf{x} = (x_1, \cdots, x_q) {\,\in\,}\mathcal{X}$, where $\mathcal{X}$ is the parameter space. A Gaussian process is a stochastic process $\{f(\mathbf{x}){\,\in\,}\mathbb{R}: \mathbf{x}{\,\in\,}\mathcal{X}\}$, for which any finite  collection of points $f(\mathbf{x}_1), \dots, f(\mathbf{x}_n)$ have a joint Gaussian distribution. A GP is fully characterized by a mean function $\mu(\mathbf{x}) = \mathbb{E}[f(\mathbf{x})]$ and a covariance function $k(\mathbf{x},\mathbf{x}') = \text{Cov}[f(\mathbf{x}),f(\mathbf{x}')]$. This will be denoted as
\[ f(\cdot) \sim \text{GP}\{\mu(\cdot),k(\cdot,\cdot)\}.\]
The mean function $\mu(\mathbf{x})$ denotes the mean of the process while the covariance function controls the smoothness of its sample paths. 

From a Bayesian perspective, the GP model $f(\cdot)$ prior to conditioning on data from the full model runs represents a modeler's prior belief on the simulation output before observing it. In practice, the mean function $\mu(\cdot)$ prior to conditioning is typically set to be a constant $\mu$. There are several popular choices for the covariance function $k(\cdot,\cdot)$, including Gaussian\footnote{%
    In the statistical literature the Gaussian function is often called a ``squared-exponential", indicated here by the superscript SE.}, Mat\'ern, and cubic covariances \cite{Rasmussen2004}.
In this study, we employ the anisotropic Gaussian covariance function, widely used for computer experiment emulators \citep{santner2003design}:
\begin{equation}
    k^{\rm SE}(\mathbf{x},\mathbf{x}') = \sigma^2 
    \exp\biggl[-\sum_{j=1}^q\frac{(x_j-x_j')^2}{2 l_j^2} \biggr].
\end{equation}
Here $\sigma^2 > 0$ is a variance parameter controlling the variation of the process around its mean, while the parameters $l_j > 0\ (j=1,2,\cdots,q)$ are characteristic length-scales. Larger $l_j$ induce stronger correlations between nearby points, resulting in smoother sample paths, whereas smaller $l_j$ result in more wiggly sample paths.

We now integrate the data obtained from the full model simulations. Suppose noisy outputs $\mathbf{y} = (y_1, \dots, y_n)$ are simulated at parameters $\mathbf{x}_1, \dots, \mathbf{x}_n$ via the sampling model
\begin{equation}
y_i = f(\mathbf{x}_i) + \epsilon_i, \quad \epsilon_i \distas{i.i.d.} N(0,\gamma^2),
\end{equation}
where $\epsilon_i$ represents statistical uncertainty, $i.i.d.$ stands for ``independent and identically distributed", and $N(0,\gamma^2)$ denotes a Gaussian normal distribution with zero mean and variance $\gamma^2$. Conditioning on the data $\mathbf{y}$ (and assuming fixed parameters $\mu$, $\sigma^2$ and $l$), the posterior distribution of $f$ at a new point on the parameter space $\mathbf{x}_{\rm new}$ can be shown to be \cite{santner2003design}
\begin{equation}
    [f(\mathbf{x}_{\rm new})|\mathbf{y}] \sim  N(\mu^*(\mathbf{x}_{\rm new}), {\sigma^2}^*(\mathbf{x}_{\rm new})),
    \label{eq:gppost}
\end{equation}
where the posterior mean and variance are given by
\begin{align}
\begin{split}
    \mu^*(\mathbf{x}_{\rm new}) &= \mu + \mathbf{k}_{\rm new}^\top (\mathbf{K}+\gamma^2 \mathbf{I}_n)^{-1} (\bm{y}-\mu\mathbf{1}_n)\\
    {\sigma^2}^*(\mathbf{x}_{\rm  new}) &= k(\mathbf{x}_{\rm new},\mathbf{x}_{\rm new}) - \mathbf{k}_{\rm new}^\top (\mathbf{K}+\gamma^2 \mathbf{I}_n)^{-1} \mathbf{k}_{\rm new}.
    \label{eq:gpeqns}
\end{split}
\end{align}
Here, $\mathbf{k}_{\rm new} = [k(\mathbf{x}_{\rm new},\mathbf{x}_i)]_{i=1}^n$ is the covariance vector between the $n$ existing design points of full-model runs and a new, interpolated point in the parameter space, and $\mathbf{K} = {[k(\mathbf{x}_i,\mathbf{x}_j)_{i,j=1}^n}]$ is the covariance matrix for the simulated data. Equations~(\ref{eq:gppost},\ref{eq:gpeqns}) provide the basis for emulator modeling: the posterior mean $\mu^*(\mathbf{x}_{\rm new})$ serves as the emulator model prediction at a new point $\mathbf{x}_{\rm new}$, and the posterior variance ${\sigma^2}^*(\mathbf{x}_{\rm  new})$ yields a quantification of emulator model uncertainty. A key appeal of GP emulators is that both their prediction and uncertainty can be efficiently computed via such closed-form expressions. In practice, the parameters $\mu$, $\sigma^2$ and $l$ are first estimated using the maximum likelihood method \cite{casella2021statistical}, then plugged into the predictive equations \eqref{eq:gpeqns} for emulation (see \cite{santner2003design} for further details on plug-in predictors).

%%%%%%%%%%%%%%%%%%%%%%%%%%%%%%%%%%%%%%%%%%%%%%
\vspace*{-2mm}
\subsection{Emulator model specification}
\label{sec2c}
\vspace*{-2mm}
%%%%%%%%%%%%%%%%%%%%%%%%%%%%%%%%%%%%%%%%%%%%%%

We now extend the above GP modeling framework to build a transfer learning emulator model. Let $f_T(\mathbf{x})$ denote the simulator output at parameter $\mathbf{x}$ for the \textit{target} system, \emph{i.e.}, the system for which data\footnote{%
    Here and in the following ``data" is short for ``full-model simulation predictions".}
are limited and emulation is desired. Let $f_S(\mathbf{x})$ denote the simulator output at parameter $\mathbf{x}$ for the \textit{source} system, \emph{i.e.}, the system for which a large set of simulation data is available. We assume that the source and target systems share the same parameter space.

We adopt the following transfer learning model linking the source and target systems:
\begin{equation}
    f_T(\mathbf{x}) = \rho f_S(\mathbf{x}) + \delta(\mathbf{x}).
\label{eq:eqst}
\end{equation}
Here, $\rho$ is a linear correlation coefficient linking the source system to the target and will be estimated from data using maximum likelihood methods. The function $\delta(\mathbf{x})$ models the \textit{discrepancy} (\emph{i.e.} systematic differences) between source and target after accounting for correlations. Since neither $f_S(\mathbf{x})$ nor $\delta(\mathbf{x})$ are known with certainty, we then place independent priors on both terms:
\begin{equation}
    f_S(\mathbf{x}) \sim \text{GP}\{\mu_S, k^{\rm SE}_S(\cdot,\cdot)\}, \quad \delta(\mathbf{x}) \sim \text{GP}\{\mu_{\delta}, k^{\rm SE}_\delta(\cdot,\cdot)\},
\label{eq:tgppriors}
\end{equation}
where different variance and length-scale parameters are used for the squared-exponential kernels $k^{\rm SE}_S$ and $k^{\rm SE}_\delta$. As before, the GP mean parameters $\mu_S$ and $\mu_\delta$, variances $\sigma^2_S$ and $\sigma^2_\delta$, and length-scales $l_S$ and $l_\delta$ are estimated from data using maximum likelihood methods.

Consider now the simulation data for training: for the source system, suppose noisy outputs $\mathbf{y}_S = \bigl(y_1^S, \dots, y_m^S\bigr)$ are available at parameters $\mathbf{X}_S=(\mathbf{x}_1^S, \dots, \mathbf{x}_m^S)$ via the sampling model
\begin{equation}
    y_i^S = f_S(\mathbf{x}_i^S) + \epsilon_i^S, \quad \epsilon_i^S \distas{i.i.d.} N(0,\gamma_S^2), \quad i=1,\dots,m.
\end{equation}
For the target system, suppose also that noisy outputs $\mathbf{y}_T = \bigl(y_1^T, \dots, y_n^T\bigr)$ are simulated at parameters $\mathbf{X}_T=(\mathbf{x}_1^T, \dots, \mathbf{x}_n^T)$ via
\begin{equation}
    y_j^T = f_T(\mathbf{x}_j^T) + \epsilon_j^T, \quad \epsilon_j^T \distas{i.i.d.} N(0,\gamma_T^2), \quad j=1,\dots,n.
\end{equation}
The goal is to to realize computational savings by keeping the sample size $n$ for the target system much smaller than the sample size $m$ for the source system.

Conditioning on both sets of data $\mathbf{y}_S$ and $\mathbf{y}_T$ (and assuming fixed GP model parameters), the posterior distribution for the target system $f_T$ at a new parameter $\mathbf{x}_{\rm new}$ can be shown to be
\begin{equation}
    [f_T(\mathbf{x}_{\rm new})|\mathbf{y}_S,\mathbf{y}_T] \sim  N(\mu^*_T(\mathbf{x}_{\rm new}), {\sigma^2_T}^*(\mathbf{x}_{\rm new})),
    \label{eq:tgppost}
\end{equation}
where the posterior mean and variance of the transfer learning emulator model are given by 
\small
\begin{align}
\begin{split}
    \mu^*_T(\mathbf{x}_{\rm new}) &= \rho\mu_S+\mu_\delta
\\
    & \quad +\mathbf{k}_{\rm new}^\top\mathbf{\Sigma}^{-1} 
    \left(
  \begin{bmatrix}
    \mathbf{y}_S\\
    \mathbf{y}_T
  \end{bmatrix} - 
  \begin{bmatrix}
    \mu_S\mathbf{1}_{m}\\
    (\rho\mu_S+\mu_\delta)\mathbf{1}_{n}
  \end{bmatrix} \right),
\\
    {\sigma^2_T}^*(\mathbf{x}_{\rm  new}) &= \rho^2\mathbf{k}_S(\mathbf{x}_{\rm new},\mathbf{x}_{\rm new})+\mathbf{k}_\delta(\mathbf{x}_{\rm new},\mathbf{x}_{\rm new}) 
\\
    & \quad - \mathbf{k}_{\rm new}^\top \mathbf{\Sigma}^{-1}\mathbf{k}_{\rm new},
\label{eq:tgpeqns}
\end{split}
\end{align}
\normalsize
with $\mathbf{k}_{\rm new} = [\mathbf{k}_{\rm new}^S,\mathbf{k}_{\rm new}^T]$ and $\mathbf{k}_{\rm new}^S = [k(\mathbf{x}_{\rm new},\mathbf{x}_i)]_{i=1}^{m}$, $\mathbf{k}_{\rm new}^T = [k(\mathbf{x}_{\rm new},\mathbf{x}_j)]_{j=1}^{n}$, and
\begin{align*}
    \mathbf{\Sigma} &= 
    \begin{bmatrix}
        \mathbf{K}_S(\mathbf{X}_S)+\gamma^2_S\mathbf{I}_{m} & \rho\mathbf{K}_S(\mathbf{X}_S,\mathbf{X}_T)\\
        \rho\mathbf{K}_S(\mathbf{X}_S,\mathbf{X}_T)^T & \rho^2\mathbf{K}_S(\mathbf{X}_T)+\mathbf{K}_\delta(\mathbf{X}_T)+\gamma^2_T\mathbf{I}_{n}
    \end{bmatrix}.
\end{align*}
Equation \eqref{eq:tgpeqns} provides the predictive equations for our transfer learning emulator model: $\mu^*_T(\mathbf{x}_{\rm new})$ serves as the emulator model prediction while ${\sigma^2_T}^*(\mathbf{x}_{\rm  new})$ quantifies its uncertainty. These closed-form equations enable efficient probabilistic predictions from the proposed model. As before, the parameters $\mu$, $\sigma^2$, $l$ and $\rho$ are estimated using maximum likelihood \cite{casella2021statistical} (first for the source, then for the discrepancy), then used in the predictive equations \eqref{eq:tgpeqns} for emulation of the target system.

The discrepancy function $\delta(\mathbf{x})$, which captures the systematic differences between the source and target, can then be estimated from equation \eqref{eq:eqst} as:
\begin{equation}
\hat{\delta}(\mathbf{x}) = \mu_T^*(\mathbf{x}) - \rho \mu_S^*(\mathbf{x}),
\label{eq:disc}
\end{equation}
where $\mu_T^*(\mathbf{x})$ is the posterior mean in equation \eqref{eq:tgpeqns} and $\mu_S^*(\mathbf{x})$ is the posterior mean of Gaussian process emulator in equation \eqref{eq:gpeqns}. A careful analysis of the estimated discrepancy function $\hat{\delta}(\mathbf{x})$ can yield useful insights on the different physics between the source and target systems. We explore this further in Section \ref{sec:sens}. 

The above transfer learning emulator model is closely related to the KO model which is widely used for multi-fidelity emulation. The KO model aims to emulate a high-fidelity computer simulation, using data simulated from lower-fidelity approximations of the \textit{same} system. The KO model is similar in spirit to Equation \eqref{eq:eqst} in that the high-fidelity code is modeled as a linear autoregressive formulation of the low-fidelity code, plus a discrepancy term to account for systematic bias. The key difference for the proposed model is that instead of transferring learning from simulations of different fidelities for the \textit{same} system, our emulator model is trained by transferring knowledge between high-fidelity simulations of \textit{different} systems that have common traits.

%%%%%%%%%%%%%%%%%%%%%%%%%%%%%%%%%%%%%%%%%%%%%%%%%%%%%%%%%%%%%%%%%%%%%%
%%%%%%%%%%%%%%%%%%%%%%%%%%%%%%%%%%%%%%%%%%%%%%%%%%%%%%%%%%%%%%%%%%%%%%
\section{Transfer Learning emulators for relativistic heavy ion collisions}
\label{sec3}
%%%%%%%%%%%%%%%%%%%%%%%%%%%%%%%%%%%%%%%%%%%%%%%%%%%%%%%%%%%%%%%%%%%%%%
%%%%%%%%%%%%%%%%%%%%%%%%%%%%%%%%%%%%%%%%%%%%%%%%%%%%%%%%%%%%%%%%%%%%%%

%%%%%%%%%%%%%%%%%%%%%%%% Table I %%%%%%%%%%%%%%%%%%%%%%%%%%%%%%%%%%%%
%\renewcommand{\arraystretch}{1.5}
\begin{table*}[!t]
\begin{tabularx}{0.8\textwidth}{|X|c|c|}
 \hline
   \multirow{2}{8em}{Observable Type}  & \multicolumn{2}{|c|}{Centralities} \\
  &  Au+Au at 0.2 TeV & Pb+Pb at 2.76 TeV\\
 \hline
Charged particle multiplicity; $dN_{ch} / d\eta$   & None   & [0-5], [60-70]\\
Pion multiplicity; $dN_{\pi}/{dy}$ &[0-5], [40-50] & [0-5], [60-70] \\
Mean transverse momenta of pions; $\langle p_{T} \rangle_{\pi}$ &[0-5], [40-50] & [0-5], [60-70]\\
Two-particle elliptic flow; $v_2\{2\}$ &   [0-5], [40-50]  & [0-5], [60-70] \\
Fluctuation in the mean transverse momentum; $\delta p_T / p_T $ &   None  & [0-5], [55-60] \\
\hline
\end{tabularx}
\caption{Observables used for emulation}
\label{tab:observables}
\end{table*}
%%%%%%%%%%%%%%%%%%%%%% end Table I %%%%%%%%%%%%%%%%%%%%%%%%%%%%%%%%%%%

All large scale Bayesian parameter estimations for relativistic heavy ion collisions have been made computationally feasible by using 
GPs as surrogates for computationally expensive simulations. The biggest computational cost associated with any such analysis is in generating training data for the GPs. In this section, we compare the accuracy and the computational cost associated with two distinct emulation methods: direct emulation with traditional Gaussian Processes and our novel transfer learning emulation method. We show that transfer learning requires significantly fewer training data from the computationally expensive simulation and thus lowers the computational barrier associated with Bayesian parameter estimation for complex problems, such as the one posed by the dynamical modeling of relativistic heavy ion collisions. Transfer learning is a particularly powerful tool for situations where (i) the training data on the target alone are insufficient to fit a good emulator, and (ii) the amount of training data available on the source is much larger than that for the target.

%%%%%%%%%%%%%%%%%%%%%%%%%%%%%%%%%%%
%%%%%%%%%%%%%%%%%%%%%%%%%%%%%%%%%%%
\subsection{Multistage model of relativistic heavy ion collision simulations}
\label{sec3a}
%%%%%%%%%%%%%%%%%%%%%%%%%%%%%%%%%%%
%%%%%%%%%%%%%%%%%%%%%%%%%%%%%%%%%%%

The relativistic heavy ion collision model used in the present work \cite{JETSCAPE:2020mzn} involves the following modules describing different evolution stages:

\renewcommand{\arraystretch}{1}

%\cmtJF{Add references to the physics papers, not just the codes.}
\begin{enumerate}
    \item \trento{}: A phenomenological model of the initial energy deposition after the impact of the nuclei \cite{Moreland:2014oya, trento_code}.
    \item Freestreaming: A model for weakly-coupled pre-equilibrium dynamics, covering the first fm/$c$ or so \cite{Liu:2015nwa, Broniowski:2008qk,fs_code}.
    \item Relativistic viscous hydrodynamics, describing the dissipative evolution of near-equilibrium QCD matter with the code \music{} \cite{Schenke:2010nt, Schenke:2010rr, Paquet:2015lta, 2000JCoPh.160..241K, hydro_code}.
    \item Particlization: Conversion of the fluid into particles after it cools down below the critical temperature where QGP converts back into hadrons, described by the Cooper-Frye formula \cite{Cooper:1974mv, Cooper:1974qi}. To parameterize the local hadron phase space distributions using only the ten components of the energy momentum tensor evolved by the hydrodynamic model, three different models with different physics assumptions are explored:
    \begin{enumerate}
        \item Grad viscous corrections, which expand the distribution function up to second order in hadron momenta \cite{Grad}; 
        \item Chapman-Enskog (CE) viscous corrections, which solve the Relaxation-Time-Approximation Boltzmann equation for linearized corrections to the distribution function \cite{chapman1990mathematical}; and
        \item Pratt-Torrieri-Bernhard (PTB) modified equilibrium viscous corrections \cite{Pratt:2010jt} which uses an exponential ansatz ensuring a positive definite distribution function. 
    \end{enumerate}
    These corrections are implemented using the \is{} sampler \cite{McNelis:2019auj, is3d_code}.
    \item Hadronic decays and re-scatterings are modeled with Boltzmann kinetic transport using the code \SMASH{} \cite{Weil:2016zrk, smash_code}. 
\end{enumerate}

To apply and test transfer learning techniques in this setting, we use a very large set of full-model simulation data that were generated for calibrating the JETSCAPE modeling framework \cite{JETSCAPE:2020mzn}, including the following systems:
\begin{enumerate}
    \item Pb+Pb collisions at $\sqrt{s_\mathrm{NN}}=2.76$\,TeV with 
    \begin{enumerate}
        \item Grad viscous corrections,
        \item Chapman-Enskog viscous corrections, and
        \item Pratt-Torrieri-Bernhard viscous corrections;
    \end{enumerate}
    \item Au+Au collisions at $\sqrt{s_\mathrm{NN}} = 0.2$\,TeV center of mass energy with Grad viscous corrections.
\end{enumerate}

All these simulations share the same set of 17 model parameters described in \cite{JETSCAPE:2020shq, JETSCAPE:2020mzn}. For model calibration, full-model simulations were performed at 500 design points that uniformly cover the 17-dimensional parameter space within a finite 17-dimensional cube described in \cite{JETSCAPE:2020shq, JETSCAPE:2020mzn}, using maximin Latin Hypercube sampling \cite{MORRIS1995381}.\footnote{%
    For technical reasons, only the simulation results from 473 of these 500 design points were used in the present analysis.
    }
For each design point and each particlization model, 2500 simulations were performed with stochastically fluctuating initial conditions and particlization results. For each design point and particlization model, a multitude of experimental observables were computed and compared with the corresponding experimental data. We use full-model predictions for only a subset of these observables (listed in Table~\ref{tab:observables}) to illustrate the proposed transfer learning emulator. For simplicity, we focus here on only two collision centralities, ``central'' ([0\%-5\%] centrality) and ``peripheral" ([40\%-50\%] centrality for the Au+Au collisions at RHIC, and [55\%-60\%] or [60\%-70\%] (whichever was the most peripheral bin available) for the Pb+Pb collisions at the LHC), and also leave out the yields and mean transverse momenta of kaons and protons, charged hadron triangular flow and transverse energy ($E_T$) distributions.  

For each choice of collision system and particlization model, we thus have a set of 473 samples of the parameter space (design points) that provide mean values and errors for each observable to train its emulator. To test the performance of the trained emulator we also generated additional test data sets for each model: 100 design points from a separate maximin Latin Hypercube design. Note that the emulators are not trained directly on the observables (as predicted by the simulations) listed in Table~\ref{tab:observables}: we first perform a standardization of each of the observables using the means and variances of the source simulation data. These transformations are slightly different from those used in~\cite{JETSCAPE:2020mzn} -- see  Appendix~\ref{app:transformation} for details.

The test data set for each model is used to evaluate the performance of each emulator by calculating the mean squared error (MSE): 
\begin{equation}
\label{eq:mse}
    \mathrm{MSE} =\sum_{
    \substack{i \in  \{\textrm{test design}\} 
    \\ 
    l \in  \{\textrm{observables}\} }}
    \frac{\left[ \hat{Y}_\mathrm{sim}^{l}(\mathbf{x_i})-\hat{Y}_\mathrm{emu}^{l}(\mathbf{x_i}) \right]^2}
{N_\mathrm{test} N_\mathrm{obs}},
\end{equation}
where $\mathbf{x_i}$ are the model parameters for the $i^{th}$ test design point and $\hat{Y}_\mathrm{sim}^l, \hat{Y}_\mathrm{emu}^l$ represent standardized (See Appendix~\ref{app:transformation})  simulation and emulation outputs for the $l^{th}$ observable. We will show plots of the MSE for target emulators constructed with $n$ target training points ($1 \leq n \leq 473$), using either the standard GP training protocol or the transfer learning protocol, and compare their performance as a function of $n$. As discussed in Sec.~\ref{sec2c}, the transfer learning emulator is trained by using these $n$ sets of target data on top of a source emulator that has been previously trained with a larger number $m$ of design points from the source system (here $m=473$).

%%%%%%%%%%%%%%%%%%%%%%%%%%%%%%%%%%%%%%%
%%%%%%%%%%%%%%%%%%%%%%%%%%%%%%%%%%%%%%%
\subsection{Transfer learning between different collision systems}
\label{sec3b}
\vspace*{-3mm}
%%%%%%%%%%%%%%%%%%%%%%%%%%%%%%%%%%%%%%%
%%%%%%%%%%%%%%%%%%%%%%%%%%%%%%%%%%%%%%%

\begin{figure}[!h]
\centering
\includegraphics[width=0.9\linewidth]{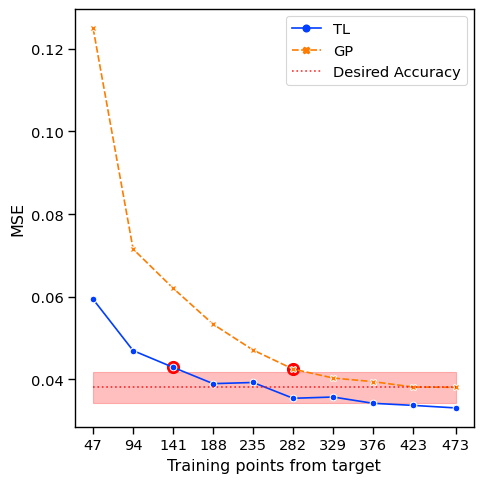}
\caption{Mean squared error prediction accuracy of emulators for Au+Au collisions at $\sqrt{s_\mathrm{NN}}=200$\,GeV using the Grad particlization model. The transfer learning emulator uses a source emulator trained on model simulations for Pb+Pb collisions at $\sqrt{s_\mathrm{NN}}=2700$\,GeV. The MSE shown is averaged over all observables, but the curves for the MSE of individual observables look all very similar. See text for discussion.
}
\label{fig:AuAuGrad}
\end{figure}

As our first application of transfer learning methods, we build emulators for simulated Au+Au collisions at $\sqrt{s_\mathrm{NN}} = 0.2$\,TeV as the target system, using available trained emulators for Pb+Pb collisions at $\sqrt{s_\mathrm{NN}} = 2.76$\,TeV as our source. The two emulation methods discussed previously are trained for each of the six observables shown in the Au+Au column of Table~\ref{tab:observables}, as a function of the number of design points $n$ for which full-model simulations of the target system are available. We do this by first randomly dividing the total set of  $n_\mathrm{max}=473$ simulation data for the target from previous work \cite{JETSCAPE:2020mzn, JETSCAPE:2020shq} into 10 roughly equal size sets (nine batches of 47 plus one batch of 50 design points). We then train the emulators using only one batch of target design points, and then repeat the training procedure after successively adding the remaining batches. After each training step, we compare the predictions for the observables from the trained emulators with the full-model test data for the 100 parameter sets in the test design, and compute its mean squared error (MSE, Eq.~(\ref{eq:mse})). The result is shown in Fig.~\ref{fig:AuAuGrad} as a function of the number $n$ of target designs used for training.

%%%%%%%%%%%%%%%%%%%%%%%%% Fig. 1 %%%%%%%%%%%%%%%%%%%%%%%%%%%%%%%%%%
%%%%%%%%%%%%%%%%%%%%%%%%%%%%%%%%%%%%%%%%%%%%%%%%%%%%%%%%%%%%%%%%%%%

The dashed orange line in the figure (labeled GP) shows the MSE for the GP emulator of the target system using the standard training protocol, without any help from the source system emulator. The dotted red horizontal line shows the final MSE reached by this method using all 473 available target design points from the full-model simulation data, with the shaded band representing a 10\% variation around this value. The solid blue line (labeled TL) shows the MSE for the proposed transfer learning emulation method, which, in addition to the $n$ target design points, also makes use of the information from the previously trained, costly GP emulator for the source system. The two red dots indicate the smallest number $n$ of target design points needed, for each emulator, for its MSE to come within 10\% of the ``asymptotic precision'' (defined by the MSE at the maximally available number of target training points) shown by the dotted red line.

The solid blue curve denoting the transfer learning MSE clearly shows that the TL emulator is more accurate than the traditional GP emulator (dashed orange curve), for all values $n$ of the number of target design points used. The relative advantage of the transfer learning emulator is particularly evident for small numbers of target system design points. For example, when using only $47$ design points for Au+Au, the transfer learning emulator has approximately half the mean squared error of the traditional emulator. Note that, even in the ``asymptotic limit" when all 473 target design points are used, the proposed transfer learning emulator still yields improved precision over the standard GP emulator, by leveraging information from the source system emulator.

As expected, for both emulator models, the emulation prediction error (in terms of MSE) decreases monotonically with increasing number of target training points $n$. For the proposed transfer learning emulator, the rate of decrease is not always uniform, which suggests that there is a diminishing marginal decrease in MSE for each additional target design point. In other words, at a certain point, the ``new" information provided by the target training data is minor compared with the ``old" information already contributed by the source system emulator.

Another way to quantify the success of the proposed transfer learning emulator is via the two large red dots in Fig.~\ref{fig:AuAuGrad}, where it can be seen that the same Au+Au collision simulation can be emulated with the same accuracy at half the number of full-model simulations. This level of success of transfer learning is quite encouraging, considering that the target here (Au+Au at $\sqrt{s_\mathrm{NN}}=200$\,GeV) involves collisions at more than an order of magnitude lower center of mass energy than the source system (Pb+Pb collisions at $\sqrt{s_\mathrm{NN}}=2760$\,GeV).

%%%%%%%%%%%%%%%%%%%%%%%%%%%%%%%%%%%%%%%
%%%%%%%%%%%%%%%%%%%%%%%%%%%%%%%%%%%%%%%
\vspace*{-2mm}
\subsection{Transfer learning between different viscous corrections at particlization}
\label{sec3c}
\vspace*{-3mm}
%%%%%%%%%%%%%%%%%%%%%%%%%%%%%%%%%%%%%%%
%%%%%%%%%%%%%%%%%%%%%%%%%%%%%%%%%%%%%%%

As discussed in Section~\ref{sec3a}, the multistage dynamical modelling of heavy ion collisions requires approximations and switching between different physical pictures which is associated with theoretical uncertainty: different modelling choices can be made in each collision stage, based on different assumptions or approximations of the governing physics. Different choices lead to models whose predictions differ from each other in quantitative detail but share qualitative features and common trends under variation of certain experimental control parameters, such as collision energy, collision centrality, system size etc. For each such model variant, teaching these trends to an emulator for its observables requires evaluating the full model at a large number of design points. Transfer learning offers a more computationally efficient strategy: after having spent large numerical resources on the training of sufficiently accurate emulators for the observables predicted for one such model variant (the source), equally accurate emulators for other variants (the targets) can be obtained at a fraction of the cost by transferring some of the qualitative tendencies from source to the targets.

%%%%%%%%%%%%%%%%%%%%%%%%%%%%%% Fig. 2 %%%%%%%%%%%%%%%%%%%%%%%%%%%%%
\begin{figure}[!ht]
\centering
\includegraphics[width=0.9\linewidth]{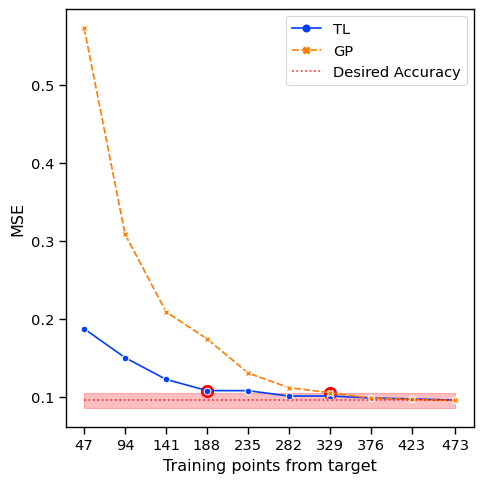}
\caption{Mean squared error prediction accuracy of emulators for Pb+Pb collisions at $\sqrt{s_\mathrm{NN}}=2760$\,GeV using the Pratt-Torrieri-Bernhard particlization model. The transfer learning emulator uses a source emulator trained on model simulations for Pb+Pb collisions at the same $\sqrt{s_\mathrm{NN}}$ using the Grad particlization model. The MSE shown is averaged over all observables, but the curves for the MSE of individual observables look all very similar. See text for discussion.
\vspace*{-3mm}
}
\label{fig:PbPbPTB}
\end{figure}
%%%%%%%%%%%%%%%%%%%%%%%%%%%%%%%%%%%%%%%%%%%%%%%%%%%%%%%%%%%%%%%%%%%

\begin{figure}[t]
\centering
\includegraphics[width=0.95\linewidth]{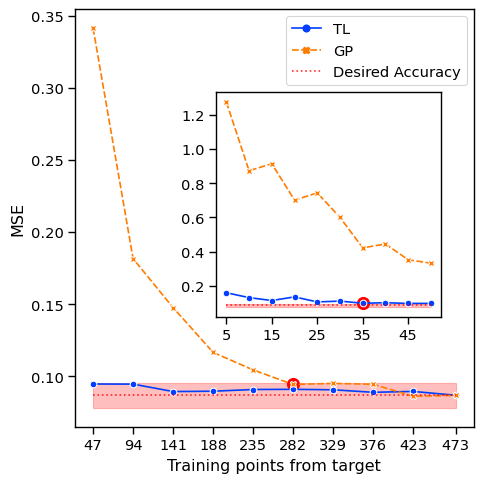}
\caption{Same as Fig.~\ref{fig:PbPbPTB} but for a target using Chapman-Enskog particlization.
\vspace*{-3mm}}
\label{fig:PbPbCE}
\end{figure}

We illustrate this idea here by considering as source and targets model variants obtained by swapping out one particular module in the multistage model, the particlization module (we refer to the discussion in Sec.~\ref{sec3a}). We consider Pb+Pb collisions at the LHC, simulated with Grad model particlization, as our ``source", and the same collisions simulated with Pratt-Torrieri-Bernhard (PTB, Fig.~\ref{fig:PbPbPTB}) or Chapman-Enskog particlization (CE, Fig.~\ref{fig:PbPbCE}) as our ``targets". 

The data we work with are the simulated model outputs for each of the three particlization models from the same design points discussed in the preceding subsection, a maximum of 473 points for emulator training plus a fixed number of 100 design points for emulator testing. Different from before, a larger set of observables is available for Pb+Pb collisions at the LHC than we had to emulate for Au+Au collisions at RHIC (c.f. Table~\ref{tab:observables}). We follow the same training strategy as described in the preceding subsection, for emulators predicting the model outputs for this larger set of observables but using the same design point batches as considered before. To zero in on the relative performance of the TL and GP emulators for the hypothetical case where only very small numbers of target model design points are available, we additionally divided the 473 total target design points to which had access randomly into smaller batches of 5 design points each, allowing studies of the evolution of the emulators'\ MSE with $n$ for smaller $n$-values (see inset in Fig.~\ref{fig:PbPbCE}). 

%%%%%%%%%%%%%%%%%%%%%%% Fig. 3 %%%%%%%%%%%%%%%%%%%%%%%%%%%%%%%%%%%%%%
%%%%%%%%%%%%%%%%%%%%%%%%%%%%%%%%%%%%%%%%%%%%%%%%%%%%%%%%%%%%%%%%%%%%%

In Figs.~\ref{fig:PbPbPTB} and \ref{fig:PbPbCE} we note that the transfer learning emulators again already approach their asymptotic accuracy within 10\% for a much smaller number of target design points than those generated with the standard GP training protocol, similar to the preceding subsection. We also note that for the case of different particlization routines shown in Figs.~\ref{fig:PbPbPTB} and \ref{fig:PbPbCE} the accuracy advantage of the TL emulators over their GP siblings begins to disappear once about 60\% of the maximally available number of target training points ($n_\mathrm{mx}=473$) have been used. 

For small numbers of target design points $n\sim 50$, the TL emulators have approximately one third or less of the mean squared error of the traditional GP emulators for the targets involving a change of particlization model, compared to the factor two reduction for the target involving a lower collision energy studied in Sec.~\ref{sec3b}. Amazingly, the inset in Fig.~\ref{fig:PbPbCE} shows that for the CE particlization model the MSE prediction accuracy of the TL emulator needs only 35 target training points to reach within 10\% of its asymptotic value, and is not much worse even for as few as only 5 target training points. This means that the qualitative trends of the observables predicted by the Grad and CE particlization models must be very close (much closer than between the source and the other two target models studied in this work), and teaching these trends to the target emulator via transfer learning almost completely obviates the need for additional information from full-model simulations of the target model. While this is clearly a special situation, it illustrates the huge cost-saving potential of transfer learning if ways can be found to reliably diagnose the convergence of the emulator accuracy towards its asymptotic value. 

%%%%%%%%%%%%%%%%%%%%%%%%%%%%%%%%%%%%%%%%%%%%%%%%%%%%%%%%%%%%%%%%
%%%%%%%%%%%%%%%%%%%%%%%%%%%%%%%%%%%%%%%%%%%%%%%%%%%%%%%%%%%%%%%%
\vspace*{-3mm}
\section{Sensitivity analysis}
\label{sec:sens}
\vspace*{-2mm}
%%%%%%%%%%%%%%%%%%%%%%%%%%%%%%%%%%%%%%%%%%%%%%%%%%%%%%%%%%%%%%%%
%%%%%%%%%%%%%%%%%%%%%%%%%%%%%%%%%%%%%%%%%%%%%%%%%%%%%%%%%%%%%%%%

There is evident interest in understanding the effect of individual model parameters on specific observables, to gain intuition about what the experimental data might tell us about the underlying physics and medium properties. This relation between parameters and observables is often explored through ``sensitivity analysis'', though the exact method varies. Examples from the field of heavy ion physics can be found in Refs.~\cite{Sangaline:2015isa, JETSCAPE:2020mzn, parkkila2021new, Everett_thesis}. 

Transfer learning offers an interesting new way of performing sensitivity analysis, by systematically investigating which model parameters contribute to non-trivial differences in parameter dependencies between source and target models. As described in Sec.~\ref{sec:transfer_learning}, Eq.~\eqref{eq:disc}, these differences can be characterized by the correlation coefficient $\rho$ and its corresponding discrepancy function $\delta(\mathbf{x})$. By estimating both $\rho$ and $\hat\delta(x)$ from data, we can then perform a sensitivity analysis on the \textit{estimated} discrepancy function $\hat \delta(\mathbf{x})$. Below, we perform such an analysis using the proposed transfer learning emulator and the scenarios discussed in the preceding section.

There are two main types of sensitivity analysis methods from the uncertainty quantification literature \cite{iooss2015review}: local or global ones. Local sensitivity analysis can quantify the model sensitivity for an observable \emph{at a fixed parameter value}, such as the maximum a posteriori (MAP) estimate obtained from parameter inference. On the other hand, global sensitivity analysis provides an averaged quantification of sensitivity for each parameter over the full parameter space. In what follows, we focus on the latter global sensitivity analysis of the estimated discrepancy function $\hat{\delta}(\mathbf{x})$ (\ref{eq:disc}).

We first introduce the first-order Sobol' indices \cite{jacques2006sensitivity}, a popular method for analyzing global sensitivity. Sobol' indices \cite{sobol1990sensitivity, im1993sensitivity} quantify the importance of each parameter for a given function $\delta(\mathbf{x})$, by decomposing its contribution to the variance of $\delta(\cdot)$ over the parameter space. The first-order Sobol' index for model parameter $x_j$ is defined as:
\begin{equation}
    \frac{\text{Var}_{X_j}(\mathbb{E}_{\bm{X}_{-j}}(\delta(\bm{X})|X_j))}
               {\text{Var}_{\bm{X}}(\delta(\bm{X}))},\qquad j=1,\dots,q.
               \label{eq:sobind}
\end{equation}
Here, $X_j$ is an independent uniform random variable for parameter $x_j$ over its parameter range, and $\bm{X} = ({X}_1, \cdots, {X}_q)$ is its corresponding random vector for all parameters. The term $\mathbb{E}_{\bm{X}_{-j}}(\delta(\bm{X})|x_j)$ is called the \textit{main effect} of parameter $x_j$: given fixed $j$-th parameter $X_j = x_j$, it \textit{averages} the function $\delta(\cdot)$ uniformly over the remaining parameters $\bm{X}_{-j} = \bm{X} \setminus X_j$. This is formally defined as
\begin{align}
\begin{split}
    &\mathbb{E}_{\bm{X}_{-j}}(\delta(\bm{X})|X_j) = 
\\
    & \quad \int_{\mathcal{X}_{-j}} \delta(x_1,\dots,x_q) \; dU(x_1,\dots,x_{j-1},x_{j+1},\dots,x_q),
\end{split}
\label{eq:me}
\end{align}
where $U(x_1,\dots,x_{j-1},x_{j+1},\dots,x_q)$ is the uniform probability measure over $\mathcal{X}_{-j}$, the parameter space $\mathcal{X}$ omitting the $j$-th parameter. The first-order Sobol' index \eqref{eq:sobind} thus quantifies the importance of parameter $x_j$, by taking the ratio of $\text{Var}_{X_j}(\mathbb{E}_{\bm{X}_{-j}}(\delta(\bm{X})|X_j))$, the variance accounted for by the \textit{main effects} $\mathbb{E}_{\bm{X}_{-j}}(\delta(\bm{X})|X_j)$, over $\text{Var}_{\bm{X}}(Y)$, the \textit{total} variance of $\delta(\cdot)$ over all parameters. For costly simulations such as for heavy ion collisions, the integral in \eqref{eq:me} can be expensive to evaluate. A standard approach \cite{iooss2015review} (which we adopt) is to replace the expensive $\delta(\cdot)$ with the estimated discrepancy $\hat{\delta}(\cdot)$ \eqref{eq:disc} from the emulator model.

One can further modify the Sobol' indices in \eqref{eq:sobind} by grouping together similar model input parameters. The grouped Sobol' indices in \cite{jacques2006sensitivity} accomplish this. The $q$ input parameters $\mathbf{X} = (X_1,\cdots,X_q)$ (assumed again to be uniformly distributed) are first divided into $J$ groups $(\mathbb{X}_1,\cdots,\mathbb{X}_J)$, given by:
$$
(X_1,\cdots,X_q) = (\underbrace{X_1,\dots,X_{k_1}}_{\mathbb{X}_1},\dots,\underbrace{X_{k_{J-1}+1},\dots,X_q}_{\mathbb{X}_J}).
$$
The first-order \textit{grouped} Sobol' indices can then be defined as:
\begin{align}
\begin{split}
    S_j = \frac{\text{Var}_{\mathbb{X}_j}(\mathbb{E}_{\mathbb{X}_{-j}}(Y|\mathbb{X}_j))}
               {\text{Var}_{\mathbf{X}}(Y)},
    \quad j=1,\cdots,J,
\end{split}
\end{align}
where $\mathbb{X}_{-j} = \mathbf{X} \setminus \mathbb{X}_j$ consists of all parameters except for those in group $j$.

In our implementation, all simulation models consider the same $q{\,=\,}17$ input model parameters. We group these parameters into six groups according to similarities of their functionality in our model. We employ the following parameter grouping:\footnote{%
    Note that the results from grouped sensitivity analysis may depend on both the grouping of parameters as well as the choice of model parameterization (\textit{e.g.}, how $\eta/s(T)$ is parameterized), thus one must be careful about the interpretation of such analyses. Further details on this can be found in \cite{borgonovo2014transformations}.}
\begin{itemize}
    \item N: The normalization parameter in \trento{}
    \item \tre{}: All other parameters in the \trento{} initial-state module.
    \item Free-streaming: Parameters controlling the free-streaming time
    \item $\eta/s$: All model inputs that parameterize the temperature dependence of the specific shear viscosity.
    \item $\zeta/s$: All model inputs that parameterize the temperature dependence of the specific bulk viscosity.
    \item $T_{sw}$: The particlization temperature separating hydrodynamics and hadronic transport.
\end{itemize}
This grouping provides meaningful insight on the global sensitivity of the discrepancy between the source and target systems. Our grouped sensitivity analysis agrees with previous sensitivity studies, while providing more concise results with clearer implications.

%%%%%%%%%%%%%%%%%%%%%%%%%%% Fig. 4 %%%%%%%%%%%%%%%%%%%%%%%%%%%%%%%%%
\begin{figure}[t]
\centering
\includegraphics[width=0.45\textwidth]{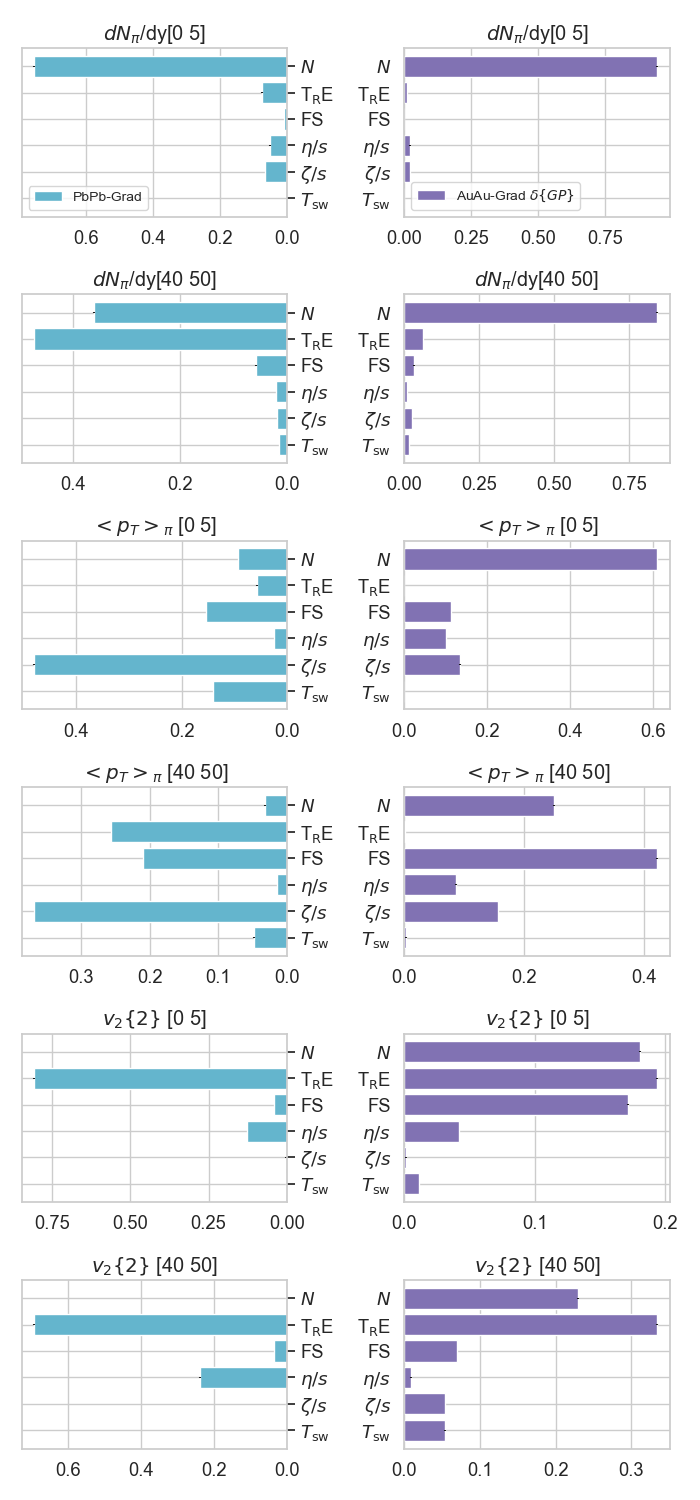}
\caption{First order group Sobol' sensitivities of the Pb+Pb 2.76 TeV source simulation (left) and of the discrepancy GP for Au+Au 200 GeV target simulation (right).
}
\label{fig:Ausensitivity}
\end{figure}
%%%%%%%%%%%%%%%%%%%%%%%%%%%%%%%%%%%%%%%%%%%%%%%%%%%%%%%%%%%%%%%%%%%%

The left column of Fig.~\ref{fig:Ausensitivity} shows the global sensitivity of the model for Pb+Pb collisions at 2.76 TeV with Grad viscous corrections, obtained from the source model emulators discussed before. The six panels in that column correspond to three different observables, each at two different centralities. Within each panel, each of the six bars represents a different group of model parameters. In central collisions (0-5\% centrality), the overall pion yield is mostly sensitive to the normalization constant $N$ for the initial energy density profile, the pion mean transverse momentum reacts most strongly to changes in the specific bulk viscosity, and the charged hadron elliptic flow is most sensitive to \trento{} model parameters (in particular, to the granularity of the initial energy density fluctuations). At first it may seem surprising that $v_2$ reacts more strongly to the \trento{} parameters than to the specific shear viscosity but this becomes clearer once one remembers that $\eta/s$ controls the {\it hydrodynamic response} to the initial-state source eccentricity $\epsilon_2$, i.e. the ratio $v_2/\epsilon_2$. The large sensitivity of $v_2$ to the \trento{} parameters really reflects their dominant effect on $\epsilon_2$ which is bigger than that of $\eta/s$ on the ratio $v_2/\epsilon_2$. In peripheral collisions, on the other hand, the left column of Fig.~\ref{fig:Ausensitivity} exhibits additional sensitivities that are much less prominent in central collisions: The overall pion yield now also exhibits sensitivity to the \trento{} parameters; this would be consistent with a stronger viscous heating effects caused by increased granularity in the smaller fireballs generated when the nuclei hit each other at larger impact parameters. The pion mean transverse momentum shows additional sensitivity to the \trento{} parameters and free-streaming time which control the early build-up of radial flow \cite{Liu:2015nwa}. And the influence of $\eta/s$ on the charged hadron $v_2$ grows in relative importance.

In the right column of Fig.~\ref{fig:Ausensitivity} we show the sensitivity of the discrepancy GPs between Pb+Pb $\sqrt{s_\mathrm{NN}}=$2.76 TeV Grad (source) and Au+Au $\sqrt{s_\mathrm{NN}}=$200 GeV (target) model outputs. Clearly, for all three observables, at both collision centralities, the discrepancy GPs share a high sensitivity to the normalization parameter $N$. This is expected since the most striking difference between these two collision systems is their total multiplicity, driven by the much higher collision energy at the LHC compared to RHIC. We further observe that the discrepancy GPs related to mean transverse momentum ($\langle p_{T} \rangle_{\pi}$) and flow observables ($v_2\{2\}$) have a significant sensitivity to the model parameters related to the pre-equilibrium stage, both via the \trento{} initialization model and the duration of the free-streaming stage. This indicates that the pre-equilibrium dynamics depends sensitively on the center of mass energy of the collision. Interestingly, the discrepancy GPs for the mean transverse momentum observable are found to be insensitive to the \trento{} parameters and the switching temperature (which is mostly constrained by the chemical composition of the final hadronic stage \cite{JETSCAPE:2020mzn}). Similarly, the discrepancy GPs for the elliptic flow observables show only weak sensitivity to the specific viscosities.\footnote{%
    The non-vanishing (albeit weak) sensitivity of $v_2\{2\}$ to the parameters describing the temperature dependence of the specific shear viscosity in central collisions and to the specific bulk viscosity in peripheral collisions supports the frequently made assertion that collisions at different center of mass energies should help to constraint the temperature dependence of these viscosities.}
In other words, these observables share roughly the same degree of sensitivity to these parameters at both collision energies -- these are the types of systematic trends in the simulations that make transfer learning efficient.

%%%%%%%%%%%%%%%%%%%%%%%%%%%% Fig. 5 %%%%%%%%%%%%%%%%%%%%%%%%%%%%%%%%%
\begin{figure}
\centering
\includegraphics[width=0.428\textwidth]{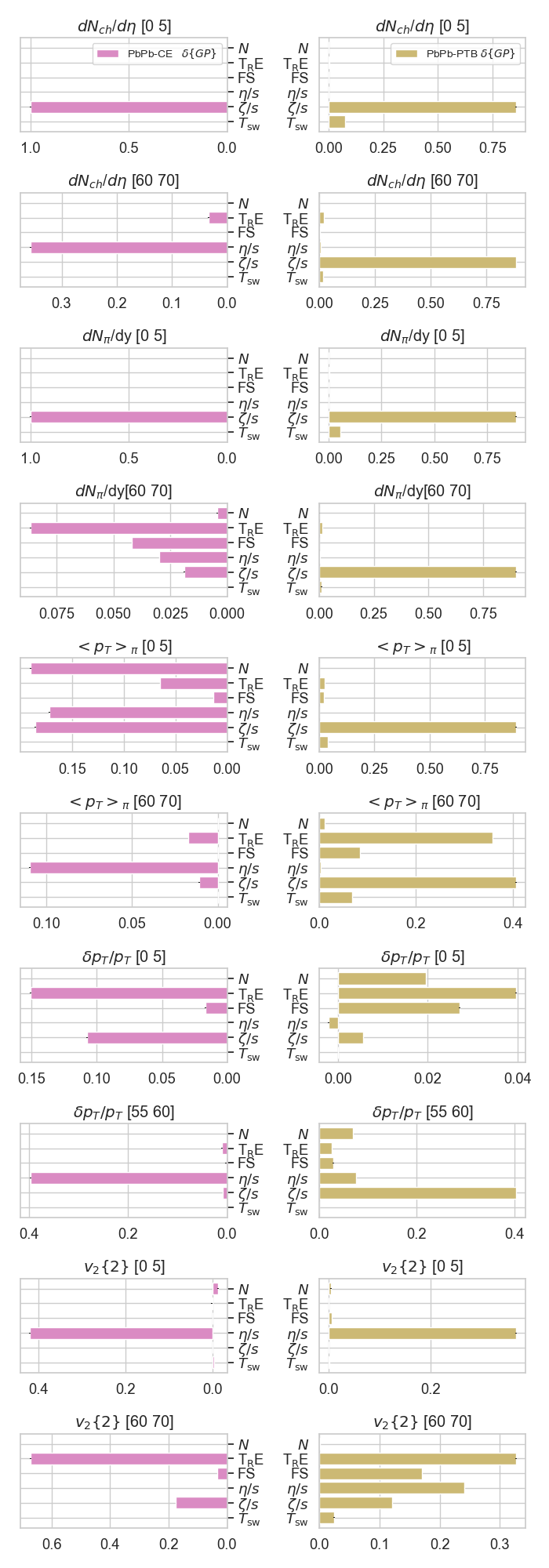}
\caption{First order group Sobol' sensitivities of discrepancy GPs. Pb+Pb with CE viscous corrections (left) and Pb+Pb with PTB viscous corrections (right).}
\label{fig:Pbsensitivity}
\end{figure}
%%%%%%%%%%%%%%%%%%%%%%%%%%%%%%%%%%%%%%%%%%%%%%%%%%%%%%%%%%%%%%%%%%%%

In Fig.~\ref{fig:Pbsensitivity} we show the analogous sensitivity plots for the discrepancy GPs for the Pb+Pb CE (left column) and Pb+Pb PTB (right column) target models.\footnote{%
    The source sensitivities shown in the left column of Fig.~\ref{fig:Ausensitivity} are the same for all three targets.}
Compared to Fig.~\ref{fig:Ausensitivity} we include two additional observables (the total charged hadron multiplicity density $dN_\mathrm{ch}/d\eta$ and the normalized $p_T$-fluctuations $\delta p_T/\langle p_T\rangle$), again for two collision centralities, resulting in ten panels for each target model. In central collisions, for both targets the majority of the discrepancy GPs (with the exception of the ones emulating the $p_T$ fluctuations and elliptic flow) are found to be most sensitive to the bulk viscosity parameters. Remembering that here the difference between source and targets is how the viscous corrections are handled during particlization, the sensitivity to the viscosity parameterizations is not surprising. More insightful is the observation that the sensitivity to the bulk viscosity sector is mostly stronger than to the shear sector. This may be related to the fact that particlization at $T_\mathrm{sw}$ happens just after hadronization of the QGP, and that the bulk viscosity peaks near the hadronization phase transition. The situation is, however, more complex in peripheral collisions where the sensitivities to the bulk and shear viscous sectors of parameter space differ between the CE and PTB targets. Furthermore, the mean values and fluctuations of the pion transverse momenta show dominant  sensitivities to different sectors of the parameter space than the other observables. All this suggests that Bayesian inference based on the available experimental data should allow us to discriminate between the different particlization models based on their ability to describe the full spectrum of observations, and that combining the strengths and weaknesses of these different models in the future via Bayesian Model Mixing \cite{Coleman_thesis, Phillips:2020dmw} may lead to overall tighter constraints on the fireball properties.

We close this section by noting that relating the source and target model emulators in the form \eqref{eq:eqst} and identifying the corresponding linear correlation coefficient $\rho$ and discrepancy $\hat \delta(\mathbf{x})$ may be a very broadly applicable technique for gaining valuable insights into qualitative similarities and differences between different models and into their success and/or failure in describing a given set of experimental data. 

\vspace*{-2mm}
%%%%%%%%%%%%%%%%%%%%%%%%%%%%%%%%%%%%%
%%%%%%%%%%%%%%%%%%%%%%%%%%%%%%%%%%%%%
\section{Computational savings from transfer learning}
\label{sec5}
%%%%%%%%%%%%%%%%%%%%%%%%%%%%%%%%%%%%%
%%%%%%%%%%%%%%%%%%%%%%%%%%%%%%%%%%%%%
\vspace*{-3mm}

Relativistic heavy ion collision experiments produce measurements for hundreds of observables. Since their dynamics is too complex to be described analytically, they are studied theoretically by building phenomenological models that are calibrated with the experimental data.
The models have multiple parameters describing properties of the collision dynamics that can not (yet) be computed from first principles and must be inferred using the experimental measurements. After calibration the models can be tested by predicting and measuring additional observables. Since both the experimental data and simulation model outputs have uncertainties associated with them, model calibration (a.k.a. solving ``the inverse problem") requires a probabilistic framework. 

As already briefly summarized in the Introduction, Bayesian parameter inference is a framework that allows for a systematic probabilistic accounting for our knowledge about the model and its uncertainties. It is based on Bayes theorem,
\begin{equation}
\label{eq:bayes}
    \mathcal{P}(\mathbf{x|y_{\rm exp}}) =  \frac{\mathcal{P}(\mathbf{y_{\rm exp}|x})\mathcal{P}(\mathbf{x})} 
         {\mathcal{P}(\mathbf{y_{\rm exp}})}.
\end{equation}
Here $\mathcal{P}(\mathbf{x})$ is prior probability for the parameters $\mathbf{x}$, and $\mathcal{P}(\mathbf{y_{\rm exp}|x})$ is the likelihood function, describing the probability that model output with a given set of model parameters $x$ agrees with the experimental data $\mathbf{y_{\rm exp}}$. It is usually assumed to be a Gaussian,
\begin{equation}
    \label{eq:likelihood}
    \mathcal{P}(\mathbf{y_{\rm exp}|x}) = \frac{1}{\sqrt{|2\pi\mathbf{\Sigma}|}}
    \exp\Bigl[-\frac{1}{2}\mathbf{y}^\top \mathbf{\Sigma}^{-1} \mathbf{y}\Bigr],
\end{equation}
where $\mathbf{y} \equiv [\mathbf{y}_\mathrm{sim}(\mathbf{x}) - \mathbf{y}_\mathrm{exp}]$ is the deviation between model prediction and experimental measurement, and $\mathbf \Sigma$ is the total uncertainty, obtained by adding the experimental and simulation uncertainties: $\mathbf{\Sigma = \Sigma_{\rm exp} + \Sigma_{\rm sim}(\mathbf{x})}$. For heavy ion collisions $\mathbf{y}$ is a vector that can have more than 100 components, and $\mathbf \Sigma$ is a quadratic matrix of the same dimensionality; $|\mathbf \Sigma|$ denotes its determinant.

The term $\mathcal{P}(\mathbf{x|y_{\rm exp}})$ on the left hand side of Eq.~\eqref{eq:bayes} is called the posterior (short for ``the posterior probability density"). It describes the probability of the model parameters $\mathbf{x}$ given the experimental data $\mathbf{y_{\rm exp}}$, and it is the main quantity of interest in Bayesian parameter inference. Its functional form is generally not known analytically, in particular not for heavy ion collisions. To find the most likely range for the parameters $\mathbf{x}$ and quantify their uncertainty requires numerical techniques for finely sampling the posterior in the neighborhood of the MAP values of the parameters. This is typically achieved by using Markov Chain Monte Carlo (MCMC) techniques.\footnote{%
    These techniques require only relative probabilities, so the normalization $\mathcal{P}(\mathbf{y_{\rm exp}})$ in the denominator on the right of Eq.~\eqref{eq:bayes} (which is independent of the parameters to be inferred) does not need to be calculated.}

For each MCMC sample of the posterior \eqref{eq:bayes} the likelihood function \eqref{eq:likelihood} must be evaluated; this requires knowledge of the model prediction $\mathbf{y}_\mathrm{sim}$ at the sampled parameter set $\mathbf{x}$. In a high-dimensional parameter space millions of MCMC samples are needed to explore the posterior in sufficient detail. In principle, this requires running the full-model simulation millions of times. For heavy ion collisions this is practically infeasible, due to the computational cost of each model simulation. This is where numerically cheap surrogate models (emulators) for $\mathbf{y}_\mathrm{sim}(\mathbf{x})$ come to the rescue. They can be trained by using \emph{very} much smaller numbers of full-model simulations (typically hundreds, not millions). They do introduce an additional emulation (or interpolation) uncertainty which is known and can be simply added to the total simulation uncertainty $\mathbf \Sigma_{\rm sim}$ when evaluating the Gaussian function \eqref{eq:likelihood}, but which we want to keep at or below the other uncertainties.

The biggest computational cost is now associated with training the emulators, which requires generating full-model simulation output at the training points. The number of training points needed to build an accurate emulator is therefore of crucial importance. For example, one of the very recent Bayesian inference attempts in relativistic heavy ion collisions \cite{parkkila2021new} which went beyond the work in \cite{JETSCAPE:2020mzn} by emulating additional observables and multiple collision systems, used 64 million CPU hours for emulator training. The authors of \cite{parkkila2021new} considered only a single evolution model which does not provide access to estimating modeling uncertainties as in \cite{JETSCAPE:2020shq}.

The analysis in \cite{JETSCAPE:2020shq} calibrated each of the different model variants by using the same set of training points, thus multiplying the cost of emulator training by the number of variants. For the extended set of collision systems and higher-statistics observables studied in \cite{parkkila2021new} this would already no longer be practical. The transfer learning technique presented in this work lowers this barrier by reducing the number of training points for subsequent model variants once an accurate emulator has been trained for the first model.

%%%%%%%%%%%%%%%%%%%%%%%%% Fig. 6 %%%%%%%%%%%%%%%%%%%%%%%%%%%%%%%%%%%%%
\begin{figure}[b]
\centering
\includegraphics[width=0.95\linewidth]{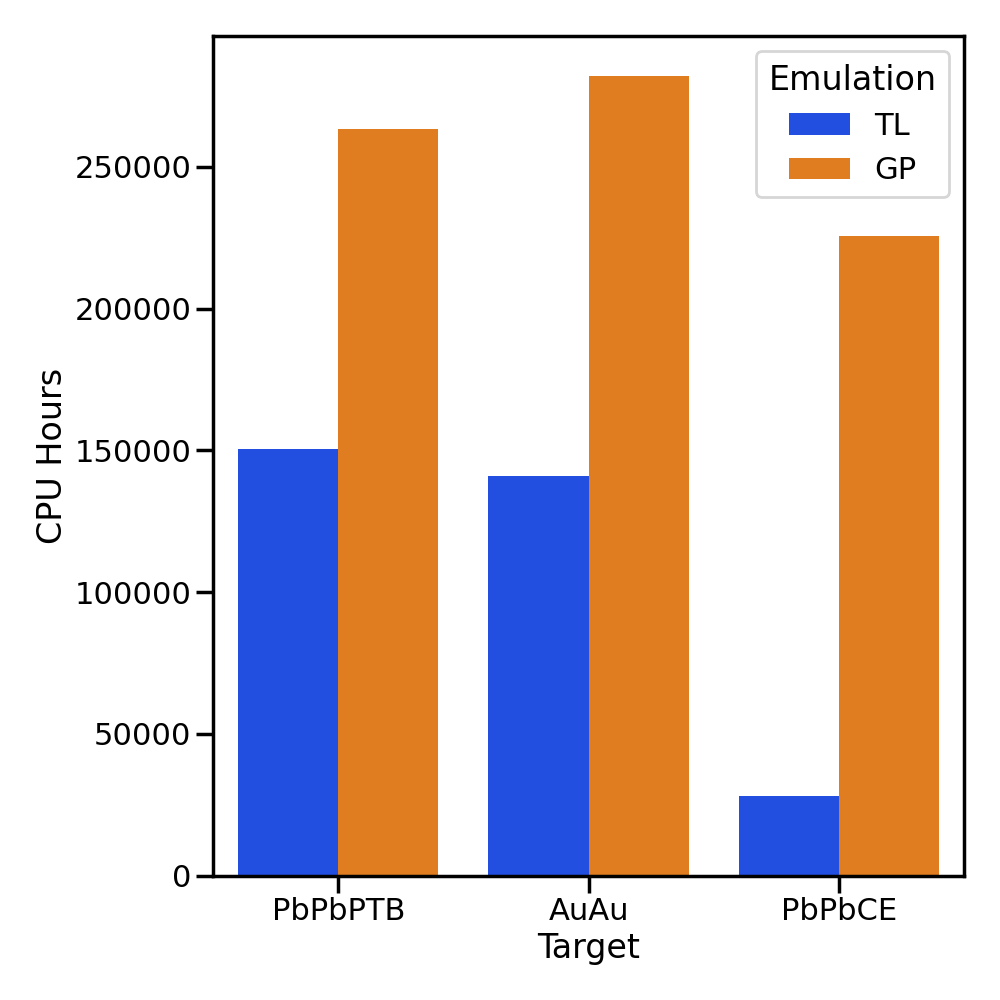}
\vspace*{-5mm}
\caption{Comparison between computational resources used by transfer learning (left blue bars) and the traditional GP emulation method (right orange bars).}
\label{fig:cpusaving}
\end{figure}
%%%%%%%%%%%%%%%%%%%%%%%%%%%%%%%%%%%%%%%%%%%%%%%%%%%%%%%%%%%%%%%%%%%%%%%

The full-model simulations used in this paper take on average $\mathcal{O}$(1000) CPU hours for each design point in model parameter space \cite{JETSCAPE:2020mzn}. A majority (80\%) of the CPU time is spent on the hadron transport stage after particlization; the remaining CPU time (20\%) is mostly utilized by the hydrodynamic QGP evolution code. In figure \ref{fig:cpusaving} we show the CPU hours needed to build accurate emulators for the three target model variants discussed in this work, with or without transfer learning from a previously trained source emulator (whose training cost was about 25\% higher than the middle orange bar).\footnote{%
    Different viscous corrections during particlization in the Pb+Pb system at LHC energies affect only the hadronic evolution after particlization. Since we take the source simulations as given, we exclude in the figure the computational cost incurred up to particlization. With this accounting, exploring the effects of different viscous corrections in the same collision system requires only $\mathcal{O}$(800) CPU hours per design point on average, for both emulation methods.}
For this plot, we decided on the required number of training samples for each emulator by requiring convergence of the mean squared error to within 10\% of the ``asymptotic" accuracy, as shown in Figs.\,\ref{fig:AuAuGrad}--\ref{fig:PbPbCE}. We note that the transfer learning method incurs significantly less computational cost compared to the standard GP training protocol. When the source and target models have a much in common (such as the Pb+Pb Grad and Pb+Pb CE models), the computational savings can exceed an order of magnitude (see right bars in Fig.~\ref{fig:cpusaving}).

We note, however, that the cost for the 100 full-model test samples needed to evaluate the MSE and the cost for determining its ``asymptotic" value are not accounted for in Fig.~\ref{fig:cpusaving}.\footnote{%
    Accounting for the cost of generating the 100 full-model test samples would add about 100,000 CPU hours to each of the bars displayed in Fig.~\ref{fig:cpusaving}.}
The (possibly large) computational cost for additional test runs can be largely avoided by using a cross-validation approach \citep{friedman2001elements}, which randomly splits the available target data into training and validation sets multiple times. One then obtains an error estimate by fitting the emulator on the training set and testing on the validation set, cycling through the different splits. Cross-validation error estimates, however, are known to be upwardly biased \citep{friedman2001elements}. This should not be a big issue when using the cross-validation MSEs as a criterion for how many full-model target simulations to use in transfer learning. For the current study, however, we were interested in a precise understanding of the convergence properties of the transfer learning method and therefore elected to use unbiased MSE estimators by running a new set of test samples for validation. 

\vspace*{-2mm}
%%%%%%%%%%%%%%%%%%%%%%%%%%%%%%%%%%%%%
%%%%%%%%%%%%%%%%%%%%%%%%%%%%%%%%%%%%%
\section{Implications for the study of heavy ion collisions}
\label{sec6}
%%%%%%%%%%%%%%%%%%%%%%%%%%%%%%%%%%%%%
%%%%%%%%%%%%%%%%%%%%%%%%%%%%%%%%%%%%%
\vspace*{-3mm}

Theoretical progress in the phenomenological study of relativistic heavy ion collisions is made by developing increasingly accurate theoretical models of the collisions that can describe both past and future experimental data. Bayesian parameter estimation in relativistic heavy ion physics approaches this aim in two different ways: First, including more experimental data in the analysis, by using multiple collision systems and adding new observables, leads to tighter bounds on the QGP properties. Second, accounting more faithfully for theoretical uncertainties results in more robust uncertainty estimates for the QGP parameters. Accounting for model differences by Bayesian Model Averaging (BMA, as done in \cite{JETSCAPE:2020shq}) usually results in weaker constraints (broader posteriors) on the plasma properties, but does not account differentially for specific strengths and weaknesses of each model in different regions of parameter space. Bayesian Model Mixing \cite{Coleman_thesis, Phillips:2020dmw} has the potential to mitigate this shortcoming, leading to modeling uncertainties that lie between those of BMA and those of a single model analysis.

For both approaches, improved knowledge extraction comes at a steep computational cost. Mitigation calls for the development of increasingly efficient emulation techniques, to reduce as much as possible the need for computationally expensive runs of increasingly complex models. This work offers transfer learning as one such instrument in the Bayesian inference tool box with the potential for significant numerical cost savings. As shown in Sec.~\ref{sec3}, it addresses both the need for including more observables and for studying multiple variants of the theoretical model. By cutting the cost of Bayesian parameter estimation, we open the door to viable systematic analyses of measurements from heavy ion data from multiple collision systems, accounting for multiple sources of theoretical model uncertainties, and yielding increasingly accurate constraints on the properties of the plasma.

%%%%%%%%%%%%%%%%%%%%%%%%%%%%%%%%%%%%%
%%%%%%%%%%%%%%%%%%%%%%%%%%%%%%%%%%%%%
\vspace*{-2mm}
\section{Conclusions and outlook}
\label{sec7}
\vspace*{-3mm}
%%%%%%%%%%%%%%%%%%%%%%%%%%%%%%%%%%%%%
%%%%%%%%%%%%%%%%%%%%%%%%%%%%%%%%%%%%%

In this work we introduced and studied transfer learning as a novel method for training emulators for relativistic heavy ion collision simulations. We showed that this method is surprisingly effective and can significantly reduce the computational cost associated with building emulators. Furthermore, we saw that there is a wealth of information in the discrepancy GP which is a by-product of transfer learning methods and offers new ways of comparison between different simulation models. To decipher the information in the discrepancy GPs, we performed a global first order Sobol' sensitivity analysis in Sec.~\ref{sec:sens}. 

The transfer learning method introduced in this work has the limitation of requiring the same set of parameters in both the target and source models. We have ideas for a more general knowledge transferring framework that can handle different parameterizations of source and target, but this will have to wait for future work.

The field of relativistic heavy ion collisions has generated a multitude of different dynamical simulation models, and their number keeps growing. A systematic approach to accurately account for the theoretical uncertainties introduced by these model ambiguities is urgently needed from a statistical and information-theoretical perspective \cite{Phillips:2020dmw}. With the present contribution we hope to help lower the barrier to implementing such a paradigm change.

%%%%%%%%%%%%%%%%%%%%%%%%%%%%%%%%%%%%%%%%%%%%%%%%%%%%%%%%%%%%%%%%%%%%%%
\vspace*{2mm}
\section*{Acknowledgments}
\vspace*{-3mm}
%%%%%%%%%%%%%%%%%%%%%%%%%%%%%%%%%%%%%%%%%%%%%%%%%%%%%%%%%%%%%%%%%%%%%%

We thank the JETSCAPE Collaboration for providing the relativistic heavy ion collision simulation data used in this work. D.L., D.E. and U.H. were supported by the NSF CSSI program under grant \rm{OAC-2004601}, and within the framework of the JETSCAPE Collaboration under NSF Award No.~\rm{ACI-1550223}, as well as by the DOE Office of Science, Office for Nuclear Physics under Award No.~\rm{DE-SC0004286}. J.-F.P. acknowledges support by DOE Award No.~\rm{DE-FG02-05ER41367}. M.H. is supported by the Natural Sciences and Engineering Research Council of Canada.
%%%%%%%%%%%%%%%%%%%%%%%%%%%%%%%%%%%%%%%%%%%%%%%%%%%%%%%%%%%%%%%%%%%%%%%%%%%%%%%%%%%%%%

%%%%%%%%%%%%%%%%%%%% APPENDIX %%%%%%%%%%%%%%%%%%%%%%%%%%%%%%%%%%%%%
\appendix
%%%%%%%%%%%%%%%%%%%%%%%%%%%%%%%%%%%%%%%%%%%%%%%%%%%%%%%%%%%%%%%%%%%

\vspace*{-2mm}
%%%%%%%%%%%%%%%%%%%%%%%%%%%%%%%%%%%%%%%%
%%%%%%%%%%%%%%%%%%%%%%%%%%%%%%%%%%%%%%%%
\section{Standardization of the observables}\label{app:transformation}
%%%%%%%%%%%%%%%%%%%%%%%%%%%%%%%%%%%%%%%%
%%%%%%%%%%%%%%%%%%%%%%%%%%%%%%%%%%%%%%%%
\vspace*{-2mm}

We standardize all simulation data before they are used to train the emulators. This is achieved by performing a standard normal transformation (\ref{eq:tf}) on the training and test data, using the means and variances of the predicted observables of our source model, i.e. for Pb+Pb collisions at $\sqrt{s_\mathrm{NN}}=2.76$\,TeV with Grad viscous corrections: 
\begin{eqnarray}
\label{eq:tf}
    &&\tilde{Y}_{j}^{l} = 
    \frac{Y_j^{l}-\mu_\textrm{Grad}^{l}}
         {\sigma_\textrm{Grad}^{l}}\,,\qquad
\\\nonumber
\label{eq:sig}
    &&\mu_\textrm{Grad}^{l} =
    \sum_{i}\frac{Y_{i,\textrm{Grad}}^{l}}
         {N_{\mathrm{train}}}\,,\ \ 
    (\sigma_\textrm{Grad}^{l})^2 
        = \sum_{i}
        \frac{\bigl(Y_{i,\textrm{Grad}}^{l}{-}\mu^{l}_{\textrm{Grad}}\bigr)^2}
             {N_{\mathrm{train}}} .
\end{eqnarray}
$Y_{i, \textrm{Grad}}^l$ is the $l^{\textrm{th}}$ observable from the source simulation $i$, and $i$ is summed over all events in the training design.

%%%%%%%%%%%%%%%%%%%%% Bibliography %%%%%%%%%%%%%%%%%%%%%
\bibliography{ref}
% \printbibliography
%%%%%%%%%%%%%%%%%%%%%%%%%%%%%%%%%%%%%%%%%%%%%%%%%%%%%%%%

\end{document}